\documentclass[journal]{IEEEtran}
\usepackage{amsmath,amssymb}
\usepackage{subcaption}
\usepackage{hyperref}
\hypersetup{hidelinks}
\hypersetup{colorlinks=true}
\usepackage{bm}
\usepackage{graphicx,graphics,color,psfrag}
\usepackage{cite,balance}
\usepackage{caption}
\usepackage{booktabs}
\usepackage{threeparttable}
\allowdisplaybreaks
\usepackage{algorithm}
\usepackage{algorithmic}
\usepackage{accents}  
\usepackage{amsthm}
\usepackage{url}
\usepackage[english]{babel}
\usepackage{multirow}
\usepackage{enumerate}
\usepackage{cases}
\usepackage{stfloats}
\usepackage{dsfont}
\usepackage{color,soul}
\usepackage{amsfonts}
\usepackage{cite,graphicx,amsmath,amssymb}
\usepackage{fancyhdr}
\usepackage{hhline}
\usepackage{graphicx,graphics}
\usepackage{array,color}
\usepackage{mathtools}
\usepackage{amsmath}
\graphicspath{{./Figs/}{./Photos/}}

\newtheorem{lemma}{\emph{\underline{Lemma}}}

\newtheorem{proposition}{\emph{\underline{Proposition}}}

\def\({\left(}
\def\){\right)}

\def\b0{{\mathbf{0}}}

\begin{document}
	\captionsetup[figure]{name={Fig.}}  
	\title{Unified Far-field and Near-field Codebook Design for XL-array Systems}
	\author{Chao Zhou,~
		Changsheng~You,~\IEEEmembership{Member,~IEEE},
		and Liujia Yao
		\thanks{All the authors are with the Department of Electronic and Electrical Engineering, Southern University of Science and Technology, Shenzhen 518055, China (e-mail: zhouchao2024@mail.sustech.edu.cn, youcs@sustech.edu.cn, yaolj2024@mail.sustech.edu.cn). 
			\emph{(Corresponding author: Changsheng You.)} \vspace{-10pt} 
		}\vspace{-20pt}} 
	
	\maketitle
	\begin{abstract}
		While extremely large-scale antenna arrays (XL-arrays) hold immense promise for next-generation wireless networks, the paradigm shift encompassing new near-field communications renders conventional discrete Fourier transform (DFT)-based Type-I codebooks inefficient. Meanwhile, existing near-field alternatives usually impose prohibitively high storage and/or feedback burdens. To tackle these challenges, we propose in this letter a \emph{unified} codebook applicable to both far-field and near-field XL-array systems. Our key idea is to express the near-field channel steering vector in an equivalent form, which is a Hadamard product of an angular vector and a \emph{curvature}-related vector. 
		This new representation allows direct reuse of the legacy Type-I codebook for angle quantization, while only a lightweight auxiliary codebook is newly added for additional curvature quantization. Subsequently, we optimize the auxiliary curvature codebook by maximizing the average beamforming gain given the information of user distribution. By leveraging the local monotonicity of the beamforming gain with respect to curvature mismatch, the optimization problem is further reduced to a quantization-error minimization problem, for which suboptimal curvature codewords are obtained. Numerical results demonstrate that the proposed unified codebook achieves near-optimal beamforming performance that approaches its upper bound based on perfect channel state information (CSI), while significantly reducing the required feedback overhead as compared to various benchmark schemes.
	\end{abstract}
	\begin{IEEEkeywords}
		Unified codebook design, near-field communications, beamforming gain.
	\end{IEEEkeywords}
	\section{Introduction}
	To meet the escalating demands of next-generation wireless networks, extremely large-scale arrays (XL-arrays) have emerged as a pivotal technology for achieving unprecedentedly high spectral efficiency and spatial resolution~\cite{YouNGAT}. By deploying thousands of antennas, the aperture of an XL-array expands significantly, rendering the users more likely to be located in the radiative near-field region of base stations (BSs). For this case, the traditional far-field planar-wavefront channel model is no longer accurate. Instead, the channels should be accurately characterized by the spherical-wavefront propagation model~\cite{LuXLMIMO,LiuPLS}. Such new channel characteristics also unlock new spatial degrees-of-freedom, enabling advanced capabilities such as near-field beam focusing and enhanced spatial multiplexing~\cite{ZhangBF,CongISAC}.
	
	Although XL-arrays offer significant advantages, they also introduce new challenges for accurate channel state information (CSI) acquisition and beamforming design.
	In practical 3rd Generation Partnership Project (3GPP) New Radio (NR) systems, codebook-based precoding standardizes data transmission by allowing the user to select a preferred candidate precoder from reference-signal measurements and report the precoding matrix indicator, rank indicator, and channel quality indicator for precoder selection at the BS~\cite{WangZC}.
	However, conventional 3GPP NR discrete Fourier transform (DFT)-based Type-I codebooks~\cite{3gpp_ts38214_v17140} rely solely on angular-domain sampling and thus suffer from considerable channel mismatch when applied to near-field spherical-wave channels.
	To address this issue, various near-field codebooks have been recently developed. For example, the polar-domain (PD) codebook was proposed in~\cite{Cui_CE} that jointly samples the angular and range domains to capture the two-dimensional sparsity of near-field channels. However, its codebook size is the product of angular and range samples, resulting in demanding storage overhead. Moreover, although the wavenumber-domain (WD) codebook was developed in~\cite{WengWD} to exploit the spatial sparsity of near-field channels, it generally incurs high feedback overhead. This is because the nonlinear mapping between the spatial and wavenumber domains requires finer codeword representations and a larger number of codewords to accurately capture channel characteristics.
	Alternatively, partition-based architectures proposed in~\cite{ChenSB} employ conventional far-field DFT codebooks over multiple subarrays to approximate near-field propagation. However, this approach generally incurs high feedback overhead because beam selection must be performed separately for each subarray, and the corresponding beam indices must be fed back individually. Moreover, the beamforming performance of subarray-based DFT codebooks usually is degraded due to the limited angular resolution in small-size arrays.
	Consequently, existing near-field codebooks face a fundamental trade-off among feedback overhead, storage overhead, and beamforming performance, making the realization of a unified design become a critical bottleneck for XL-array systems.
	
	Motivated by the above, we propose in this letter a unified codebook design for XL-array systems. To this end, we firstly show that the near-field steering vector admits a separable \emph{angle-curvature} structure. Based on this observation, we develop a unified angle-curvature codebook that allows reuse of the legacy Type-I codebook for angular-domain representation and exploits an additional lightweight auxiliary codebook for curvature quantization, hence achieving backward compatibility with legacy far-field oriented codebook designs. Since the Type-I codebook is directly reused for angle quantization, we focus on optimizing the auxiliary curvature codebook to maximize the average beamforming gain under a uniform user distribution. By exploiting the local monotonicity of the beamforming gain with respect to (w.r.t.) curvature mismatch, we reformulate the codebook design problem as a quantization error minimization problem and obtain high-quality curvature codewords that are asymptotically optimal in the high-resolution regime. Numerical results show that the proposed codebook achieves superior beamforming performance which approaches its upper bound based on perfect CSI with only two curvature-sampling bits, hence significantly reducing feedback overhead while outperforming state-of-the-art benchmarks.
	
	\vspace{-10pt}
	\section{System Model}\label{Sec:SM}
	We consider a narrowband downlink communication system, where a BS equipped with an $N$-antenna XL-array, indexed by $\mathcal{N} \triangleq \{1,2,\ldots,N\}$, serves users uniformly distributed over a considered region in the Cartesian plane, spanning both the near- and far-field regions. Without loss of generality, the XL-array is deployed along the $y$-axis with its array center located at the origin $\mathbf{u}_0 = [0,0]^T$. The coordinate of the $n$-th antenna is given by $\mathbf{u}_n = [0,u_n]^T,~\forall n\in\mathcal{N}$, where $u_n = \frac{2n-N-1}{2}d,~\forall n\in\mathcal{N}$, with $d \triangleq \lambda/2$ denoting the inter-antenna spacing and $\lambda$ being the carrier wavelength.
	\vspace{-10pt}
	\subsection{Channel Model}
	We adopt spherical wavefronts to characterize the radio propagation channels associated with XL-arrays. For an arbitrary user located in the near-field region, the multi-path channel, denoted by $\mathbf{h}_{\rm u}^{H}\in\mathbb{C}^{1 \times N}$, can be modeled as~\cite{YaoCB}
	\begin{align}\label{Eq:MPchannel}
		\mathbf{h}_{\rm u}^{H}
		= \sqrt{N} h_{0}\mathbf{a}^{H}(\theta_{\rm u}, r_{\rm u})
		+ \sqrt{\frac{N}{L}} \sum_{\ell=1}^{L} h_{\ell}\mathbf{a}^{H}(\theta_{\ell}, r_{\ell}),
	\end{align}
	which consists of one line-of-sight (LoS) path and $L$ non-LoS (NLoS) paths. Herein, $h_{0}$ and $h_{\ell}$ denote the complex-valued gains of the LoS path and the $\ell$-th NLoS path, respectively, while $(\theta_{\rm u}, r_{\rm u})$ and $(\theta_{\ell}, r_{\ell})$ represent the spatial angles and ranges w.r.t. the array center. For high-frequency systems, the NLoS components are typically much weaker than the LoS component. Hence, the channel in \eqref{Eq:MPchannel} can be approximated by its LoS component as $ \mathbf{h}_{\rm u}^{H} \approx \sqrt{N} h_{0}\mathbf{a}^{H}(\theta_{\rm u}, r_{\rm u})$, where the near-field steering vector is given by
	\begin{align}\label{eq:steering_fresnel}
		[\mathbf{a}(\theta,r)]_n
		\approx
		\frac{1}{\sqrt{N}}
		e^{\jmath\frac{2\pi}{\lambda}(u_n\theta-\frac{u_n^2(1-\theta^2)}{2r})},~\forall n\in\mathcal{N},
	\end{align}
	Herein, the approximation in~\eqref{eq:steering_fresnel} is made based on the second-order Taylor expansion $r_n = \sqrt{r^2 + u_n^2 - 2 r u_n \theta} \approx r - u_n \theta + \frac{u_n^2 (1 - \theta^2)}{2r} $~\cite{Cui_CE}, which is accurate when $r_{\rm u}\ge 0.5\sqrt{ {D^3}/{\lambda}}$ with $D=(N-1)d$ denoting the array aperture. Note that the near-field steering vector in~\eqref{eq:steering_fresnel} reduces to its far-field counterpart when the range is sufficiently large, e.g., $r_{\rm u}\ge 2D^2/\lambda$.
	
	\section{Proposed Unified Codebook Design}
	In this section, we propose a unified codebook framework applicable to both far-field and near-field communication systems. The corresponding codewords are then obtained by solving an associated codeword design problem.
	
	\vspace{-10pt}
	\subsection{Proposed Unified Codebook}\label{Sec:II-B}
	To design a unified codebook, one important observation drawn from \eqref{eq:steering_fresnel} is that the near-field steering vector exhibits a separable angle-curvature structure. Specifically, by defining the curvature parameter as $\mu \triangleq \frac{1-\theta^2}{2r}$, the channel steering vector in \eqref{eq:steering_fresnel} can be equivalently rewritten as
	\begin{align}\label{eq:decomp}
		\mathbf{a}(\theta,r)
		&=
		\frac{1}{\sqrt{N}}
		\left[
		e^{\jmath\frac{2\pi}{\lambda}u_1\theta},
		\ldots,
		e^{\jmath\frac{2\pi}{\lambda}u_N\theta}
		\right]^T
		\nonumber\\
		&\odot
		\left[
		e^{-\jmath\frac{2\pi}{\lambda}u_1^2\mu},
		\ldots,
		e^{-\jmath\frac{2\pi}{\lambda}u_N^2\mu}
		\right]^T
		\triangleq
		\mathbf{b}(\theta)\odot\mathbf{c}(\mu),
	\end{align}
	where $\odot$ denotes the Hadamard product, $\mathbf{b}(\theta)$ encapsulates the angle-dependent linear phase shift, and $\mathbf{c}(\mu)$ captures the range-dependent quadratic phase curvature.
	
	By exploiting the structural property in \eqref{eq:decomp}, we propose a unified codebook applicable to both near-field and far-field communication systems, parameterized by the angle-curvature pair $(\hat{\theta},\hat{\mu})$. Let $\mathcal{C}_{\theta}$ denote a predefined Type-I angle codebook with cardinality $|\mathcal{C}_{\theta}|=2^{p}$. To characterize the near-field phase variation, $q$ feedback bits are dedicated to quantizing the curvature parameter. Let $\mathcal{S}_{\mu}\triangleq \{\hat{\mu}_{1},\ldots,\hat{\mu}_{2^{q}}\}$ denote the discrete curvature sampling set. The corresponding curvature codebook is then constructed as
	$	\mathcal{C}_{\mu} 
	=
	\left\{
	\mathbf{c}(\hat{\mu}_{i})
	:
	\hat{\mu}_{i}\in\mathcal{S}_{\mu}
	\right\}$,
	where each curvature codeword is given by
	\begin{align}
		\mathbf{c}(\hat{\mu}_{i})
		=
		\left[
		e^{-\jmath\frac{2\pi}{\lambda}u_{1}^{2}\hat{\mu}_{i}},
		\ldots,
		e^{-\jmath\frac{2\pi}{\lambda}u_{N}^{2}\hat{\mu}_{i}}
		\right]^{T}.
	\end{align}
	Accordingly, given an angular codeword $\mathbf{b}(\hat{\theta}) \in \mathcal{C}_{\theta}$ and curvature parameter $\hat{\mu} \in \mathcal{S}_{\mu}$, the composite near-field beamforming codeword can be expressed as
	\begin{align}\label{eq:codeword_unified}
		\mathbf{w}(\hat{\theta},\hat{\mu})
		=
		\mathbf{b}(\hat{\theta})\odot \mathbf{c}(\hat{\mu}).
	\end{align}
	
	The proposed unified codebook exhibits appealing advantages in  three key aspects. 
	1)~\emph{Feedback protocol}: The framework operates using a shared codebook dictionary 
	between the BS and the users. During the CSI feedback phase, the user estimates the downlink channel, selects the optimal angle and curvature indices, and feeds them back to the BS. 
	The composite codeword then can be reconstructed via the element-wise product of two key vectors defined in \eqref{eq:codeword_unified}. This structure accommodates both far-field and near-field scenarios by leveraging the widely adopted legacy Type-I codebook in the angular domain, as well as a lightweight auxiliary curvature codebook.
	2)~\emph{Multi-path extension}: Since each path in~\eqref{Eq:MPchannel} inherently follows a separable angle-curvature representation, this codeword structure can be extended to multi-path scenarios by selecting $(L+1)$ angle-curvature pairs and additionally feeding back quantized magnitudes and phases of $\{h_\ell\}$ for coherent combination.
	3)~\emph{Feedback overhead}: The total $B$ feedback bits for a single path are partitioned into $p$ bits for the angle and $q$ bits for the curvature. Notably, the $q$ bits required for the curvature codebook are generally fewer than the $p$ bits allocated for the angle. This ensures an efficient implementation that captures near-field spherical wavefront characteristics without imposing excessive overhead compared to legacy systems.

	\vspace{-8pt}
	\subsection{Problem Formulation}
	As presented in Section~\ref{Sec:II-B}, the proposed unified codebook seamlessly reuses the widely adopted legacy Type-I codebook in the angle domain. With sufficiently many angular-feedback bits, the angular mismatch becomes negligible (i.e., $\hat{\theta}\approx \theta$), and the challenge of codebook design reduces to optimizing the auxiliary curvature codebook $\mathcal{C}_{\mu}$, or equivalently, the discrete curvature sampling set $\mathcal{S}_{\mu}$. 
	Specifically, given curvature $\mu$ and a quantized curvature parameter $\hat{\mu} \in \mathcal{S}_{\mu}$, the beamforming gain can be expressed as
	\begin{align}
		G(\mu,\hat{\mu})
		=&
		\big|
		\big(
		\mathbf{b}(\hat{\theta})\odot\mathbf{c}(\mu)
		\big)^{H}
		\big(
		\mathbf{b}(\hat{\theta})\odot\mathbf{c}(\hat{\mu})
		\big)
		\big| \nonumber\\
		=
		&\frac{1}{N}\Big|\sum_{n\in \mathcal{N}} 
		\big(
		e^{-\jmath\frac{2\pi}{\lambda}u_n\hat{\theta}}
		e^{\jmath\frac{2\pi}{\lambda}u_n^2\mu}
		\big)
		\big(
		e^{\jmath\frac{2\pi}{\lambda}u_n\hat{\theta}}
		e^{-\jmath\frac{2\pi}{\lambda}u_n^2\hat{\mu}}
		\big)\Big| \nonumber\\
		=&
		\frac{1}{N}\Big|
		\sum_{n\in \mathcal{N}} 
		e^{
			-\jmath\frac{2\pi}{\lambda}u_{n}^{2}(\mu-\hat{\mu})
		}
		\Big|\triangleq G(\Delta \mu), \label{eq:gain_exact_mu}
	\end{align}
	where $\Delta \mu\triangleq \mu-\hat{\mu}$ denotes the curvature mismatch. It is observed from \eqref{eq:gain_exact_mu} that the near-field beamforming gain depends exclusively on $\Delta \mu$.
	
	Our objective is to design the curvature sampling set $\mathcal{S}_{\mu}$ to maximize the expected beamforming gain. As such, the curvature codebook optimization problem is formulated as
	\begin{align}
		\textbf{(P1)}:\;
		\max_{\mathcal{S}_{\mu}} \quad &
		\mathbb{E}_{\mu|\hat{\theta}}
		\left[
		\max_{\hat{\mu}\in\mathcal{S}_{\mu}} G(\Delta \mu)
		\right] \nonumber\\
		\text{s.t.}\quad 
		&|\mathcal{S}_{\mu}|=2^{q}. \label{C:P1_size}
	\end{align}
	Herein, the expectation is taken over the conditional distribution of the curvature parameter $\mu$ given $\hat{\theta}$. Constraint~\eqref{C:P1_size} sets the cardinality of the curvature codebook to $2^q$, corresponding to a feedback overhead of $q$ bits.

	It is worth noting that Problem \textbf{(P1)} is generally intractable to solve directly, due to the lack of a closed-form expression for the average beamforming gain, as well as the coupled non-convex constraints in \eqref{C:P1_size}.
	\vspace{-10pt}
	\subsection{Proposed Solution to Problem~\textbf{(P1)}}	
	In this subsection, we develop an efficient approach to optimize the curvature codebook.
	\subsubsection{Problem reformulation} 
	To solve Problem \textbf{(P1)}, we first characterize the local monotonicity of the beamforming gain $G(\Delta \mu)$ w.r.t. the curvature mismatch $\Delta \mu$ as below.

	\begin{lemma}[Local monotonicity of $G(\Delta \mu)$]\label{Lem:Mono} \rm
		Under the continuous-aperture approximation, the near-field beamforming gain $G(\Delta \mu)$ is monotonically decreasing w.r.t. $|\Delta \mu|$ when
		\begin{align}
			|\Delta \mu|
			\le
			\frac{v_0^2\lambda}{D^2}
			\approx
			\frac{3.65\lambda}{D^2},
			\label{eq:fresnel_bound}
		\end{align}
		where $D=(N-1)d$ denotes the array aperture, and $v_0\approx 1.91$ is the first positive root of 
		\begin{align}
			v\!\Big[
			C(v)\cos\!\big(\tfrac{\pi}{2}v^2\big)
			+
			S(v)\sin\!\big(\tfrac{\pi}{2}v^2\big)
			\Big]
			=
			C^2(v)+S^2(v),
			\label{eq:v0_root}
		\end{align}
		with $C(v)=\int_0^v \cos\!\left(\tfrac{\pi}{2}t^2\right)\,\mathrm{d}t$ and $S(v)=\int_0^v \sin\!\left(\tfrac{\pi}{2}t^2\right)\,\mathrm{d}t$ representing the Fresnel cosine and sine integrals, respectively.
	\end{lemma}
	\begin{proof}
			Based on \eqref{eq:gain_exact_mu}, the beamforming gain $G(\Delta \mu)$ is an even function which is exclusively determined by the curvature mismatch $\Delta \mu$. As such, without loss of generality, we restrict our analysis to $\Delta \mu \ge 0$. Furthermore, when the number of antennas is sufficiently large, the beamforming gain in \eqref{eq:gain_exact_mu} can be tightly approximated by the following continuous integral
		\begin{align}
			G(\Delta \mu)
			&\approx
			\Big|
			\frac{1}{D}
			\int_{-D/2}^{D/2}
			e^{-\jmath\frac{2\pi}{\lambda}x^2\Delta \mu
			}\,\mathrm{d}x
			\Big| 
			=
			\Big|
			\frac{1}{v}
			\int_{0}^{v}
			e^{-\jmath\frac{\pi}{2}t^2}\,\mathrm{d}t
			\Big|  \nonumber\\
			&=\frac{1}{v}\sqrt{\big(\int_0^v \cos\!\left(\tfrac{\pi}{2}t^2\right)\,\mathrm{d}t\big)^2 + \big(\int_0^v \sin\!\left(\tfrac{\pi}{2}t^2\right)\,\mathrm{d}t\big)^2 }, \nonumber\\
			& = {\sqrt{C^2(v)+S^2(v)}}/{v} \triangleq G_{\rm c}(v),
		\end{align}
		where $v=D\sqrt{{\Delta \mu}/{\lambda}}$, $C(v)$ and $S(v)$ are the standard Fresnel cosine and sine integrals as defined in {\bf Lemma \ref{Lem:Mono}}.
		
		Next, to characterize the stationary points of $G_{\rm c}(v)$ in the feasible region where $G_{\rm c}(v)>0$, an efficient way is to equivalently analyze its squared value. Specifically, let $\widetilde{G}(v) \triangleq G_{\rm c}^2(v) =\frac{C^2(v)+S^2(v)}{v^2}$. By applying the chain rule, we obtain
		\begin{align}
			\frac{\mathrm{d}\widetilde{G}(v)}{\mathrm{d}v}
			=
			2G_{\rm c}(v)\frac{\mathrm{d}G_{\rm c}(v)}{\mathrm{d}v}.
		\end{align}
		Consequently, for $G_{\rm c}(v)>0$, the derivatives of $G_{\rm c}(v)$ and $\widetilde{G}(v)$ have the same sign and vanish at the same points. By taking the derivative of $\widetilde{G}(v)$ w.r.t. $v$, we obtain
		\begin{align}
			\frac{\mathrm{d}\widetilde{G}(v)}{\mathrm{d}v}
			=
			\frac{
				2v\big[C(v)C'(v)+S(v)S'(v)\big]
				-
				2\big[C^2(v)+S^2(v)\big]
			}{v^3}. \nonumber
		\end{align}
		By substituting $C'(v)=\cos\left(\tfrac{\pi}{2}v^2\right)$ and $S'(v)=\sin\left(\tfrac{\pi}{2}v^2\right)$ into $\frac{\mathrm{d}\widetilde{G}(v)}{\mathrm{d}v}=0$, we obtain \eqref{eq:v0_root}. Let $v_0$ denote the first positive root of \eqref{eq:v0_root}, which corresponds to the first stationary point of $G_{\rm c}(v)$. Since $\widetilde{G}'(v)$ is continuous, negative near zero, and nonzero on $(0,v_0)$, $G_{\rm c}(v)$ is strictly decreasing on this interval. By substituting $v=v_0$ into $v=D\sqrt{\Delta\mu/\lambda}$, we obtain
		\begin{align}
			|\Delta \mu|
			\le
			\frac{v_0^2\lambda}{D^2}
			\approx
			\frac{3.65\lambda}{D^2},
		\end{align}
		thus completing the proof.
	\end{proof}

	Based on the above monotonic interval, where larger mismatches reduce the beamforming gain, we consider enough curvature-feedback bits such that the curvature mismatch strictly remains within the local monotone region established in \textbf{Lemma~\ref{Lem:Mono}}. As such, maximizing the beamforming gain $G(\mu,\hat{\mu})$ is mathematically equivalent to minimizing the absolute curvature mismatch $|\Delta \mu|$ (i.e., $|\mu-\hat{\mu}|$). As such, Problem \textbf{(P1)} can be reformulated as minimizing the average absolute curvature quantization error, which is given by
	\begin{align}
		\textbf{(P2)}:\;
		\min_{\mathcal{S}_{\mu}} \quad 
		\mathbb{E}_{\mu|\hat{\theta}}
		\Big[
		\min_{\hat{\mu}\in\mathcal{S}_{\mu}} |\mu- \hat{\mu}|
		\Big]  \quad 
		\text{s.t.}\quad  \eqref{C:P1_size}.  \nonumber 
	\end{align}
	
	\subsubsection{Proposed solution to Problem (P2)}
	In this subsection, we develop a suboptimal solution to the non-convex optimization Problem~\textbf{(P2)} under the assumption that users are uniformly distributed in the considered region.
	
	To facilitate the resolution of Problem~\textbf{(P2)}, we first characterize the statistical distribution of the near-field curvature parameter. Specifically, let $r \in [r_{\min}, r_{\max}]$ denote the physical user distance. Given the resolved spatial angle $\hat{\theta}$, the conditional distribution of the user range $r$ is given by
	\begin{align}
		f_{R|\Theta}(r|\hat{\theta})=\frac{2r}{r_{\max}^2-r_{\min}^2}, \quad r\in[r_{\min},r_{\max}].
	\end{align}
	
	To facilitate the analysis, we define an auxiliary variable $\hat{\vartheta} \triangleq 1-\hat{\theta}^2$.
	Then, for a given $\hat{\theta}$, the near-field curvature parameter can be expressed as $\mu=\frac{\hat{\vartheta}}{2r}$, which establishes a strictly monotonic relationship between the user range $r$ and the curvature $\mu$. Accordingly, the inverse transformation is given by $r=\frac{\hat{\vartheta}}{2\mu}$, whose Jacobian is $|\frac{\mathrm{d} r}{\mathrm{d} \mu}|=\frac{\hat{\vartheta}}{2\mu^2}$. Based on this transformation, the range boundaries $r_{\min}$ and $r_{\max}$ can be equivalently mapped to the curvature domain as
	$\mu_{\min}=\frac{\hat{\vartheta}}{2r_{\max}}$ and 
	$\mu_{\max}=\frac{\hat{\vartheta}}{2r_{\min}}$, respectively. Based on the above, the conditional probability density function (PDF) of $\mu$ given $\hat{\theta}$ is obtained as
	\begin{align}
		f_{\mu|\Theta}(\mu|\hat{\theta})
		\!=\!
		f_{R|\Theta}(r|\hat{\theta})
		\left|\frac{\mathrm{d}r}{\mathrm{d}\mu}\right| =
		\frac{\hat{\vartheta}^2}{2(r_{\max}^2-r_{\min}^2)\mu^3},
	\end{align}
	for $\mu\in[\mu_{\min},\mu_{\max}]$. For notational simplicity, we define $f_\mu(\mu)\triangleq f_{\mu|\Theta}(\mu|\hat{\theta})$. Accordingly, Problem~\textbf{(P2)} can be re-expressed as the following curvature sampling optimization problem
	\begin{align}
		\textbf{(P3)}:\;
		\min_{\mathcal{S}_\mu}\quad
		\int_{\mu_{\min}}^{\mu_{\max}}
		\min_{\hat{\mu}\in\mathcal{S}_\mu}
		|\mu-\hat{\mu}|\, f_\mu(\mu)\,\mathrm{d}\mu 
		\quad 
		\text{s.t.}\quad  \eqref{C:P1_size}.  \nonumber 
	\end{align}
	
	To efficiently solve Problem~\textbf{(P3)}, we partition the feasible curvature interval $[\mu_{\min},\mu_{\max}]$ into $N_\mu\triangleq 2^q$ disjoint quantization cells,  denoted by $\{\mathcal{I}_i\}_{i=1}^{N_\mu}$.  Specifically, let $\mathcal{I}_i=[\bar\mu_{i-1},\bar\mu_i)$, where $\mu_{\min}=\bar\mu_0<\bar\mu_1<\cdots<\bar\mu_{N_\mu}=\mu_{\max}$. By defining $\hat\mu_i\in\mathcal{I}_i$ as the representative curvature sample for the $i$-th cell, Problem~\textbf{(P3)} can be equivalently reformulated as 
	\begin{subequations}\label{eq:P4_set}
		\begin{align}
			\textbf{(P4)}:\;
			\min_{\{\mathcal{I}_i\},\{\hat{\mu}_i\}}
			\quad
			&\sum_{i=1}^{N_\mu}
			\int_{\mathcal{I}_i}
			|\mu-\hat{\mu}_i|\,f_\mu(\mu)\,\mathrm{d}\mu  \nonumber \\
			\text{s.t.}\quad
			&\mathcal{I}_i \cap \mathcal{I}_j = \emptyset, \quad \forall i\neq j, \label{eq:P4_set_b}\\
			&\bigcup_{i=1}^{N_\mu}\mathcal{I}_i = [\mu_{\min},\mu_{\max}], \label{eq:P4_set_c}\\
			&\hat{\mu}_i \in \mathcal{I}_i, \quad \forall i=1,\ldots,N_\mu. \label{eq:P4_set_d}
		\end{align}
	\end{subequations}
	In the following, we propose a two-stage optimization framework to sequentially obtain the optimal representative samples and the corresponding cell boundaries.
	
	\begin{proposition}[Optimal curvature samples]\label{prop:sample_update}\rm 
		For any given partition $\{\mathcal{I}_i\}_{i=1}^{N_\mu}$, the optimal representative sample for the $i$-th cell, denoted by $\hat\mu_i^\star$, is given by
		\begin{align}\label{eq:sample_update}
			\hat\mu_i^\star = \sqrt{\frac{2\bar\mu_{i-1}^2\bar\mu_i^2}{\bar\mu_{i-1}^2+\bar\mu_i^2}}, \quad \forall i \in \{1,\ldots,N_\mu\}.
		\end{align}
		Accordingly, the corresponding minimum expected sampling error within the $i$-th cell is given by
		\begin{align}\label{eq:cell_min_error}
			D_i(\hat\mu_i^\star) \!=\! c_\mu \Big( \sqrt{2\big(\frac{1}{\bar\mu_{i-1}^2} + \frac{1}{\bar\mu_i^2}\big)} - \big(\frac{1}{\bar\mu_{i-1}} + \frac{1}{\bar\mu_i}\big) \Big),
		\end{align}
		where $c_\mu\triangleq \frac{\hat{\vartheta}^2}{2(r_{\max}^2-r_{\min}^2)}$ is  a normalization constant.
	\end{proposition}
	
	\begin{proof}
		For a given cell $\mathcal{I}_i$, let $D_i(\hat\mu_i) = \int_{\bar\mu_{i-1}}^{\bar\mu_i} |\mu-\hat\mu_i|\,f_\mu(\mu)\,\mathrm{d}\mu$ denote the expected absolute error. By splitting the integral domain at $\hat\mu_i$, we obtain
		\begin{align}
			D_i(\hat\mu_i) = \int_{\bar\mu_{i-1}}^{\hat\mu_i} (\hat\mu_i-\mu) f_\mu(\mu)\,\mathrm{d}\mu + \int_{\hat\mu_i}^{\bar\mu_i} (\mu-\hat\mu_i) f_\mu(\mu)\,\mathrm{d}\mu. \nonumber
		\end{align}
		By applying the Leibniz integral rule, we set the first-order derivative of $D_i(\hat\mu_i)$ w.r.t. $\hat\mu_i$ to zero, which yields
		\begin{align}\label{eq:leibniz}
			\int_{\bar\mu_{i-1}}^{\hat\mu_i} f_\mu(\mu)\,\mathrm{d}\mu = \int_{\hat\mu_i}^{\bar\mu_i} f_\mu(\mu)\,\mathrm{d}\mu.
		\end{align}
		By substituting $f_\mu(\mu) = c_\mu \mu^{-3}$ into \eqref{eq:leibniz} and calculating the definite integrals, we have
		\begin{align}
			-\frac{1}{2}(\hat\mu_i^\star)^{-2} + \frac{1}{2}\bar\mu_{i-1}^{-2} = -\frac{1}{2}\bar\mu_i^{-2} + \frac{1}{2}(\hat\mu_i^\star)^{-2},
		\end{align}
		leading to the closed-form expression in \eqref{eq:sample_update}. Substituting $\hat\mu_i^\star$  into the split integrals yields the result in \eqref{eq:cell_min_error}.
	\end{proof}
	
	Based on {\bf Proposition~\ref{prop:sample_update}}, the optimal cell boundaries $\{\bar\mu_i\}$ can be obtained by using iterative numerical methods, such as the Lloyd-Max algorithm, which alternates between updating boundaries and representation points. However, this iterative method lacks a closed-form solution and incurs prohibitively high computational complexity. To overcome this difficulty, we exploit the asymptotic properties of high-resolution quantization theory to obtain a near-optimal solution in closed form for the partition boundaries.
	
	\begin{proposition}[Asymptotically optimal partition boundaries]\label{prop:boundary_update}\rm
		When the codebook size $N_\mu=2^q$ is sufficiently large, the partition boundaries that minimize  Problem~\textbf{(P3)} are given by
		\begin{align}\label{eq:boundary_update}
			\!\!\bar\mu_i^\star \!=\! \big( \mu_{\min}^{-\frac{1}{2}} - \frac{i}{N_\mu} ( \mu_{\min}^{-\frac{1}{2}} - \mu_{\max}^{-\frac{1}{2}} ) \big)^{-2},\forall i \in \{0,\ldots,N_\mu\}.
		\end{align}
	\end{proposition}
	\begin{proof}
		To minimize the total quantization error $\sum_{i=1}^{N_\mu} D_i(\hat\mu_i^\star)$, we introduce an auxiliary sequence $x_i \triangleq \frac{1}{\bar\mu_i}$ for $i=0,\ldots,N_\mu$. Then, the expected error of the $i$-th cell in \eqref{eq:cell_min_error} can be rewritten as $E_i(x_{i-1}, x_i) = c_\mu \big( \sqrt{2(x_{i-1}^2 + x_i^2)} - (x_{i-1} + x_i) \big)$.  Next, we define $z_i \triangleq \sqrt{x_i} = 1/\sqrt{\bar\mu_i}$. For high resolution codebooks, adjacent boundaries are narrowly spaced. By applying the second-order Taylor expansion $\sqrt{1+y} \approx 1 + \frac{1}{2}y - \frac{1}{8}y^2$ to the square root term,  the local cell error can be approximated as $E_i \approx c_\mu \frac{(z_{i-1} - z_i)^2(z_{i-1} + z_i)^2}{4z_{i-1}^2}$. Under the high resolution assumption, we have $\frac{(z_{i-1} + z_i)^2}{4z_{i-1}^2} \approx 1$. As a result, the total distortion can be approximated as
		$\sum_{i=1}^{N_\mu} E_i \approx c_\mu \sum_{i=1}^{N_\mu} (z_{i-1} - z_i)^2$. Accordingly, the boundary optimization reduces to minimizing this quadratic sum subject to the boundary constraints $z_0 = 1/\sqrt{\mu_{\min}}$ and $z_{N_\mu} = 1/\sqrt{\mu_{\max}}$. By applying the Cauchy-Schwarz inequality, we obtain
		$	\sum_{i=1}^{N_\mu} (z_{i-1} - z_i)^2 \ge \frac{1}{N_\mu} \big( \sum_{i=1}^{N_\mu} (z_{i-1} - z_i) \big)^2 = \frac{(z_0 - z_{N_\mu})^2}{N_\mu}$. 
		The equality holds if and only if $z_{i-1} - z_i = \frac{z_0 - z_{N_\mu}}{N_\mu}, \forall i$. 
		As such, the optimal sequence $\{{z_i^\star}\}$ is given by $z_i^\star = z_0 - i \frac{z_0 - z_{N_\mu}}{N_\mu},\forall i$. By substituting $z_0$, $z_{N_\mu}$, and $z_i = 1/\sqrt{\bar\mu_i}$
		into the above expression, the optimal partition boundary $\bar{\mu}_i^\star$ is given in \eqref{eq:boundary_update}.
	\end{proof}
	
	By substituting the asymptotically optimal boundaries $\{\bar\mu_i^\star\}$ into \eqref{eq:sample_update}, the curvature codebook is constructed in closed form as $\mathcal{S}_\mu^\star = \{\hat\mu_1^\star,\ldots,\hat\mu_{N_\mu}^\star\}$, where
	\begin{align}
		\hat\mu_i^\star(\hat\theta) = \sqrt{\frac{2(\bar\mu_{i-1}^\star)^2(\bar\mu_i^\star)^2}{(\bar\mu_{i-1}^\star)^2+(\bar\mu_i^\star)^2}},~\forall i \in \{1,\ldots,N_\mu\}.
	\end{align}
	
	\section{Simulation Results}\label{Sec:SR}

	We consider an XL-array system, where the BS, operating at $f_c = 30$ GHz and equipped with $N=256$ antennas, serves users uniformly distributed over an annular sector in the Cartesian plane, bounded by an angular interval of $[-60^\circ,60^\circ]$ and a range interval of $[9,200]$ m. Unless otherwise specified, the number of angular samples is set to $N_\theta=2^{B_\theta}=1024$, corresponding to $B_\theta=10$ feedback bits, and the number of curvature samples is set to $N_\mu=2^{B_\mu}=8$, corresponding to $B_\mu=3$ feedback bits.
	Moreover, the following benchmark schemes are considered for performance comparison. 
	1)~\textbf{Perfect CSI:} The beamforming vector is designed based on the perfect CSI, serving as the upper bound. 
	2)~\textbf{Type-I codebook:} The traditional far-field DFT codebook is considered, which solely samples the angular domain using $B_\theta=13$ quantization bits. 
	3)~\textbf{PD codebook:} This refers to the near-field polar-domain codebook that jointly samples the angular and distance domains, where the angular and distance sampling bits are set to $B_\theta=10$ and $B_r=3$, respectively.
	4)~\textbf{WD codebook:} This is a wavenumber-domain codebook that represents the near-field beam by a sparse combination of wavenumber-domain basis vectors~\cite{WengWD}. The wavenumber dictionary size is $N_{\rm WD}=2^{B_{\rm WD}}=1024$, with $B_{\rm WD}=10$. Unless otherwise specified, $S_{\rm WD}=2$ active wavenumber components are selected. For each selected component, $B_\alpha=3$ and $B_\varphi=5$ bits are used to quantize its magnitude and phase, respectively, yielding $2\times(10+3+5)=36$ feedback bits in total. 
	5)~\textbf{Type-I codebook + Subarray:} A partition-based architecture is considered, where the XL-array is divided into $M_{\rm sub}=8$ subarrays, each with $N_{\rm sub}=N/M_{\rm sub}=32$ antennas~\cite{ChenSB}. A far-field Type-I codebook with $N_{\theta,{\rm sub}}=2^{B_{\theta,{\rm sub}}}=128$ angular samples is applied within each subarray, where $B_{\theta,{\rm sub}}=7$. With one subarray as the phase reference, $B_\phi=3$ bits are used to quantize each of the $M_{\rm sub}-1$ relative phases, yielding $8\times7+7\times3=77$ feedback bits in total.
	
	\begin{table}[t]
		\centering
		\caption{Comparison of feedback and storage overhead.}
		\vspace{-4pt}
		\label{tab:fb_storage}
		\renewcommand{\arraystretch}{0.8}
		\begin{threeparttable}
			\begin{tabular}{lcc}
				\toprule
				Scheme & Feedback bits & Storage overhead \\
				\midrule
				Proposed codebook
				& 
				13
				& $\mathcal{O}\!\left(N\,(N_{\theta}+N_{\mu})\right)$
				\\
				
				PD codebook
				& 
				13
				& $\mathcal{O}\!\left(N\,N_{\theta}\,N_{r}\right)$ 
				\\
				
				Type-I Codebook
				&
				13
				& $\mathcal{O}\!\left(N\,N_{\theta}\right)$
				\\
				
				Type-I Codebook + Subarray
				& 
				77
				& $\mathcal{O}\!\left(N_{\mathrm{sub}} N_{\theta,\mathrm{sub}}\right)$ 
				\\
				
				WD Codebook
				& 
				36
				& $\mathcal{O}\!\left(N\,N_{\mathrm{WD}}\right)$ 
				\\
				\bottomrule
			\end{tabular}
		\end{threeparttable}
	\vspace{-14pt}
	\end{table}
	
	First, Table~\ref{tab:fb_storage} compares the feedback and storage overheads of the considered codebook schemes. In terms of storage overhead, the PD codebook needs to store a full angle-range codebook containing $N_\theta N_r$ vectors of length $N$, resulting in a multiplicative storage overhead of $\mathcal{O}(N N_{\theta}N_{r})$. In contrast, the proposed codebook decouples the angular and curvature components, which are stored separately with a curvature dictionary shared across angles, thereby reducing the storage overhead to the additive form $\mathcal{O}(N(N_{\theta}+N_{\mu}))$. Although the Type-I codebook uses $N_\theta=2^{13}$ angular vectors with a storage overhead of $\mathcal{O}(N N_{\theta})$, it only captures the angular dimension and thus cannot accurately characterize near-field spherical wavefronts. Moreover, all subarrays share one dictionary of $N_{\theta,{\rm sub}}$ vectors of length $N_{\rm sub}$, while the WD codebook stores $N_{\rm WD}$ basis vectors of length $N$. Therefore, the proposed codebook achieves a better balance among feedback overhead, storage overhead, and beamforming performance.

\begin{figure}[t]
		\centering
		\begin{subfigure}[b]{0.46\linewidth}
			\includegraphics[width=1\linewidth]{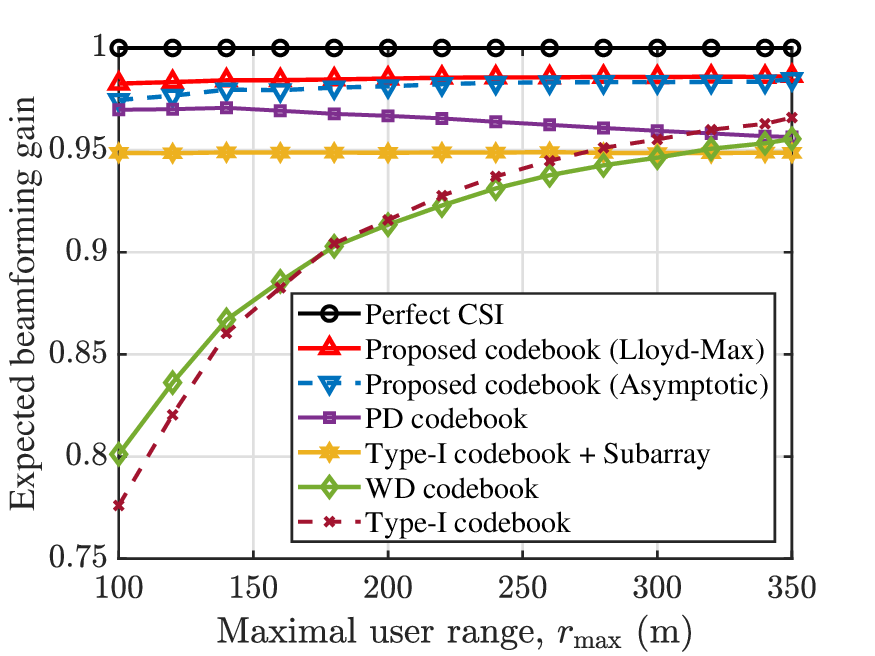}
			\caption{Maximal user range.}
			\label{Fig:BGvsRange}
		\end{subfigure}
		\hspace{2pt}
		\begin{subfigure}[b]{0.46\linewidth}
			\includegraphics[width=1\linewidth]{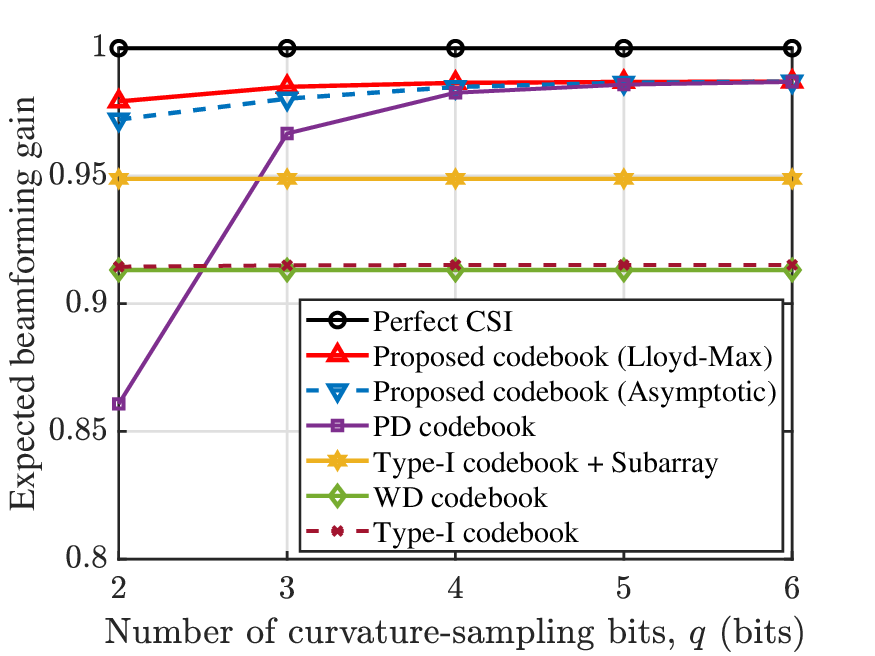}
			\caption{Curvature-sampling~bits.}
			\label{Fig:BGvsBits}
		\end{subfigure}
		\caption{Expected beamforming gain performance.}
		\label{Fig:All}
		\vspace{-14pt} 
	\end{figure}

	Next, we present the expected beamforming gain versus the maximal user range in Fig.~\ref{Fig:All}(a), where a smaller $r_{\max}$ corresponds to stronger near-field effects. As expected, the conventional Type-I codebook suffers a pronounced performance degradation when $r_{\max}$ decreases to $100$ m, because of the severe mismatch caused by the far-field assumption. The WD codebook also exhibits noticeable degradation in the small-range regime. By contrast, the proposed codebook (both Lloyd-Max and Asymptotic designs) consistently outperforms the subarray and PD schemes and achieves performance close to the perfect CSI upper bound. Moreover, it maintains a stable, remarkably high expected beamforming gain across the entire range of~$r_{\max}$.
	
	Last, Fig.~\ref{Fig:All}(b) plots the expected beamforming gain against the number of curvature-sampling bits, $q$.
	It is observed that as $q$ increases, the asymptotic performance of the proposed codebook rapidly approaches that of the Lloyd-Max design. More importantly, the proposed codebook can achieve near optimal performance, obtaining an expected beamforming gain higher than $0.95$ even with only $q=2$ bits.
	In contrast, the PD codebook is much more sensitive to the distance-quantization resolution and suffers a noticeable performance loss when $q=2$. To close the performance gap with the proposed codebook, the PD codebook requires at least $q=4$ bits.
	\vspace{-6pt} 
	\section{Conclusions}\label{Sec:Con}	

	In this letter, we proposed a unified codebook design for far- and near-field XL-array systems. By exploiting the separable angle-curvature structure of spherical-wave steering vectors, the proposed unified codebook extends the legacy Type-I codebook with a lightweight curvature codebook, reducing the storage overhead from multiplicative to additive scaling. The curvature codebook design was reformulated as a quantization problem, for which both the optimal representative samples and a low-complexity asymptotic partition were obtained. Numerical results showed that the proposed codebook achieves robust beamforming performance over different maximal user ranges and approaches the perfect-CSI benchmark with low feedback and storage overhead.

	\bibliographystyle{IEEEtran}
	\bibliography{CBdesign.bib}

\begin{thebibliography}{10}
\providecommand{\url}[1]{#1}
\csname url@samestyle\endcsname
\providecommand{\newblock}{\relax}
\providecommand{\bibinfo}[2]{#2}
\providecommand{\BIBentrySTDinterwordspacing}{\spaceskip=0pt\relax}
\providecommand{\BIBentryALTinterwordstretchfactor}{4}
\providecommand{\BIBentryALTinterwordspacing}{\spaceskip=\fontdimen2\font plus
\BIBentryALTinterwordstretchfactor\fontdimen3\font minus
  \fontdimen4\font\relax}
\providecommand{\BIBforeignlanguage}[2]{{%
\expandafter\ifx\csname l@#1\endcsname\relax
\typeout{** WARNING: IEEEtran.bst: No hyphenation pattern has been}%
\typeout{** loaded for the language `#1'. Using the pattern for}%
\typeout{** the default language instead.}%
\else
\language=\csname l@#1\endcsname
\fi
#2}}
\providecommand{\BIBdecl}{\relax}
\BIBdecl

\bibitem{YouNGAT}
C.~You \emph{et~al.}, ``Next generation advanced transceiver technologies for
  {6G} and beyond,'' \emph{{IEEE} J. Sel. Areas Commun.}, vol.~43, no.~3, pp.
  582--627, Mar. 2025.

\bibitem{LuXLMIMO}
H.~Lu \emph{et~al.}, ``A tutorial on near-field {XL-MIMO} communications toward
  {6G},'' \emph{{IEEE} Commun. Surv. Tut.}, vol.~26, no.~4, pp. 2213--2257,
  2024.

\bibitem{LiuPLS}
T.~Liu \emph{et~al.}, ``Physical-layer security in mixed near-field and
  far-field communication systems,'' \emph{IEEE Trans. Cogn. Commun. Netw.},
  vol.~12, pp. 4045--4059, 2026.

\bibitem{ZhangBF}
H.~Zhang, C.~You, and C.~Zhou, ``Near-field beam-focusing pattern under
  discrete phase shifters,'' \emph{{IEEE} Trans. Wireless Commun.}, vol.~25,
  pp. 631--646, 2026.

\bibitem{CongISAC}
J.~Cong \emph{et~al.}, ``Near-field integrated sensing and communication:
  Opportunities and challenges,'' \emph{{IEEE} Wireless Commun.}, vol.~31,
  no.~6, pp. 162--169, Dec. 2024.

\bibitem{WangZC}
Z.~Wang, H.~Wu, Y.~Chen, and L.~Lu, ``New paradigm for unified near-field and
  far-field wireless communications,'' \emph{{IEEE} Netw.}, vol.~39, no.~6, pp.
  112--118, Nov. 2025.

\bibitem{3gpp_ts38214_v17140}
\BIBentryALTinterwordspacing
{3GPP}, ``{NR; Physical layer procedures for data},'' 3rd Generation
  Partnership Project (3GPP), Technical Specification TS 38.214 V17.14.0, Jul.
  2025, release 17. [Online]. Available:
  \url{https://www.etsi.org/deliver/etsi_ts/138200_138299/138214/17.14.00_60/ts_138214v171400p.pdf}
\BIBentrySTDinterwordspacing

\bibitem{Cui_CE}
M.~Cui and L.~Dai, ``Channel estimation for extremely large-scale {MIMO}:
  Far-field or near-field?'' \emph{{IEEE} Trans. Commun.}, vol.~70, no.~4, pp.
  2663--2677, Apr. 2022.

\bibitem{WengWD}
C.~Weng, X.~Guo, Y.~Guo, and Y.~Wang, ``Wavenumber domain beam training in
  {XL-MIMO} systems: Unifying far-field and near-field,'' \emph{{IEEE} Trans.
  Wireless Commun.}, vol.~25, pp. 2183--2196, 2026.

\bibitem{ChenSB}
K.~Chen, C.~Qi, C.-X. Wang, and G.~Y. Li, ``Beam training and tracking for
  extremely large-scale {MIMO} communications,'' \emph{{IEEE} Trans. Wireless
  Commun.}, vol.~23, no.~5, pp. 5048--5062, May 2024.

\bibitem{YaoCB}
L.~Yao, C.~You, Z.~Huang, C.~Zhou, Z.~Yang, and X.~Li, ``Codebook design for
  limited feedback in near-field {XL-MIMO} systems,'' \emph{{IEEE} Trans.
  Commun.}, vol.~74, pp. 6426--6442, 2026.

\end{thebibliography}
	
\end{document}